\documentclass[letterpaper]{mn2e}
\usepackage{psfig}
\usepackage[totalwidth=510pt, totalheight=710pt]{geometry}

\catcode`\@=11
\def\gsim{\ifmmode{\,\mathrel{\mathpalette\@versim>\,}}
    \else{$\,\mathrel{\mathpalette\@versim>}\,$}\fi}
\def\lsim{\ifmmode{\,\mathrel{\mathpalette\@versim<\,}}
    \else{$\,\mathrel{\mathpalette\@versim<}\,$}\fi}
\def\@versim#1#2{\lower 2.9truept \vbox{\baselineskip 0pt \lineskip
    0.5truept \ialign{$\m@th#1\hfil##\hfil$\crcr#2\crcr\sim\crcr}}}
\catcode`\@=12  

\def\Mcltot{M_{\rm cl}} 

\def\Mstar{M_{\rm *}}

\def\Ndm{N_{\rm DM}} 
\def\Nlum{N_{\rm lum}}

\def\rhostar{\rho_*}
\def\rhodm{\rho_{\rm DM}}
 
\def\rhocl{\rho_{\rm cl}}

\def\Msol{M_{\odot}}

\def\avbeta{\langle \beta \rangle}
\def\avrdmfin{\langle \rdmfin/\rccl \rangle}

\def\Dt{\Delta t} 
\def\Dtmax{\Delta t_{\rm max}} 
\def\Dtmin{\Delta
t_{\rm min}}

\def\rccl{r_{\rm cl}} 
\def\rdm{r_{\rm DM}} 
\def\rdmfin{r_{\rm DM,f}} 
 
\def\rcstar{r_{\rm *}}

\def\sg0{\sigma_0}

\def\Td{T_{\rm dyn}}

\def\alphatol{\alpha_{\rm tol}}
\def\thetamin{\theta_{\rm min}}

\def\tout7{T_{\rm out,7}}

\def\cmm3{\,{\rm cm}^{-3}}

\def\sm1{\,{\rm s}^{-1}}

\def\kpc{{\rm kpc}}

\title[Galactic cannibalism and CDM profiles] {Galactic cannibalism and CDM density profiles}

\author[C. Nipoti, T. Treu, L. Ciotti, and M. Stiavelli]
{C.~Nipoti$^{1}$\thanks{E-mail: nipoti@thphys.ox.ac.uk}, T.~Treu$^{2,3}$, L.~Ciotti$^4$, and M.~Stiavelli$^5$\\
$^{1}$Theoretical Physics, Oxford University,  1 Keble Road, Oxford OX1 3NP, UK\\
  $^2$Department of Physics \& Astronomy, UCLA, Box 951547, Los Angeles, CA 90095-1547\\
  $^3$Hubble fellow\\
$^{4}$Department of Astronomy, Bologna University, via Ranzani 1, 40127 Bologna, Italy\\
$^5$Space Telescope Science Institute,  3700 San Martin Drive, Baltimore, MD 21218}
\begin{document}


\date{Accepted, 8th September 2004; in original form, 5th April 2004}      

\pagerange{\pageref{firstpage}--\pageref{lastpage}} \pubyear{2004}

\maketitle
\label{firstpage}

\begin{abstract}

Using N-body simulations we show that the process of formation of the
brightest cluster galaxy through dissipationless galactic cannibalism
can affect the inner cluster dark matter density profile.  In
particular, we use as realistic test case the dynamical evolution of
the galaxy cluster C0337-2522 at redshift $z=0.59$, hosting in its
centre a group of five elliptical galaxies which are likely to be the
progenitor of a central giant elliptical.  After the formation of the
brightest cluster galaxy, the inner cluster dark matter density
profile is significantly flatter (logarithmic slope $0.49 \lsim \beta
\lsim 0.90$) than the original cusp ($\beta=1$), as a consequence of
dynamical friction heating of the massive galaxies against the diffuse
cluster dark matter. In our simulations we have assumed that the
cluster galaxies are made of stars only. We also show that the
presence of galactic dark matter haloes can steepen the cluster
central density profile.  We conclude that galactic cannibalism could
be a viable physical mechanism to reconcile -- at least at the cluster
scale -- the flat dark matter haloes inferred observationally in some
galaxy clusters with the steep haloes predicted by cosmological
simulations.
\end{abstract}


\section{Introduction}

Cold Dark Matter (CDM) cosmological simulations predict that the inner
density profile of dark matter (DM) haloes is characterized by a cusp:
$\rhodm (r)~\propto~r^{-\beta}$, with logarithmic slope $\beta \sim
1-1.5$ for $r \to 0$: the exact value of the slope and its
universality are at the centre of a lively debate. For example, several
authors (Navarro, Frenk \& White 1996, NFW; Moore et al. 1998; Ghigna
et al.\ 2000; Navarro et al.\ 2004) claim that the resulting profiles
are universal, in the sense that $\beta$ is independent of the halo
mass, while other authors suggest that the inner profile depends
significantly on the slope of the power spectrum, so that the higher
is the mass of the halo the steeper is the cusp (Subramanian, Cen \&
Ostriker 2000; Ricotti 2003, and references therein).

These predictions have been extensively tested by observations both in
galaxies (e.g., Salucci \& Burkert 2000; van den Bosch et al. 2000; de
Blok et al.\ 2001; Simon et al.\ 2003) and in clusters of galaxies
(e.g., Smith et al. 2001; Sand, Treu \& Ellis 2002; Kelson et al.\
2002; Gavazzi et al. 2003; Lewis, Buote \& Stocke 2003; Sand et
al. 2004). The derived values of $\beta$ range from $\sim0$ to
$\sim1.3$. The observational results -- shallow slopes and intrinsic
scatter -- could represent a serious challenge to the standard
cosmological model, and exotic scenarios have been also proposed to
solve this problem (e.g. Spergel \& Steinhardt 2000).

In any case, the role of baryons must be better understood before we
are forced to reject a successful paradigm. In fact, it is well known
that on scales of the order of kiloparsecs baryons are important and
their evolution may affect substantially the DM distribution (see
Binney 2004 for a discussion). For example, in the so-called
``adiabatic contraction'' approximation (Blumenthal et al. 1986; 
  Gnedin et al. 2004), baryon dissipation effectively steepens the
inner density profile of the host halo (Mo, Mao \& White 1998;
Kochanek \& White 2001), thus exacerbating the contrast between theory
and observations.

Here, motivated by recent observational studies of galaxy clusters
based on a joint lensing and dynamical analysis -- in which Sand et
al.  (2002, 2004) report inner logarithmic slopes in the
range\footnote{The uncertainties on $\beta$ could be higher than those
  given by Sand et al. (2004) if the ellipticity of the cluster mass
  distribution is higher than estimated by the authors (cf. Dalal \&
  Keeton 2004; Bartelmann \& Meneghetti 2004).  Additional
  observational work is currently under way to clarify those issues
  and measure the distribution of DM inner slopes for larger samples
  of clusters.}  $\beta=0.38-0.99$ -- we focus on the cluster-sized DM
haloes, which are considered to be less affected by baryon
dissipation, with respect to galaxy-sized haloes.

We note that the spiraling in of massive galaxies in clusters could
flatten the cluster DM distribution, by heating it through dynamical
friction (similar to the galaxy-globular cluster interaction at
smaller scales; see Bertin, Liseikina \& Pegoraro 2003).  Recently,
El-Zant et al. (2004), using N-body simulations, found that dynamical
friction heating due to galaxies, modeled as rigid baryons clumps, is
indeed effective in flattening the inner DM profile in clusters.
Similar conclusions, in a different context, were also reached by Ma
\& Boylan-Kolchin (2004), who explored with N-body simulations the
effects of the dynamical evolution of (deformable) DM sub-haloes on
the DM distribution of their host halo. On the other hand, the mass
infall associated with the spiraling in of the galaxies towards the
cluster centre deepens the central potential well. This
dissipationless contraction, in contrast with dynamical friction
heating, can result in shrinking the cluster DM distribution and
steepening its inner density profile (see, e.g., Barnes \& White 1984;
Jesseit, Naab \& Burkert 2002): the final density profile will be the
result of the two competing effects (see also Section~4).

Here we investigate the effects of dynamical friction on the cluster
DM distribution in a realistic scenario, by simulating the time
evolution of the galaxy cluster C0337-2522 at redshift $z=0.59$ (ROSAT
Deep Cluster Survey; Rosati et al. 1998). This cluster is
characterized by the presence of five bright elliptical galaxies (Es)
located within $\sim30$ kpc from the centre, in projection. Nipoti et
al. (2003b, hereafter N03), using N-body simulations, showed that the
five galaxies are very likely to merge and form a brightest cluster
galaxy (BCG) by $z=0$, in accordance with the predictions of the
galactic cannibalism scenario (Ostriker \& Tremaine 1975; see also
Merritt 1983; Dubinski 1998).  In the present work we consider the
evolution of the underlying cluster DM distribution, with the aid of
additional higher resolution simulations (see also Nipoti et al.
2003c).  In our models the single galaxies, as well as the cluster DM,
are represented by N-body distributions: this is particularly
important, because tidal stripping could remove mass from the
individual galaxies, thus altering the infall process and its effects
on the cluster DM halo.

\section{Numerical simulations}

\subsection{Initial conditions}

Our simulations represent possible realizations of the dynamical
evolution of C0337-2522 -- a poor galaxy cluster, with five bright Es
in its centre -- from redshift $z=0.59$ to the present. The
observational data and the set up of the initial conditions are
described in detail in N03. 

In the initial conditions of the simulations the cluster is
represented as a live spherically symmetric DM distribution, hosting
five identical massive galaxies modeled as spherical $N$-body systems.
For simplicity, we only consider simulations where each galaxy is
represented by a single collisionless component (stars).  This is
justified under the assumption that the mass of the galaxies in the
cluster centre is dominated by luminous matter (e.g. Treu \& Koopmans
2004), and that the extended DM halo is likely to be tidally stripped
in the innermost region of the cluster (e.g. Natarajan, Kneib \& Smail
2002).

\begin{table}
  \caption{Parameters of the simulations.}
\begin{tabular}{lrrrrrr} \#
  &$\rccl$&$\Mcltot$&$\Ndm$&$\Nlum$&$\rdmfin/\rccl$&$\beta \qquad$  \\
\hline
1  &  100     & 4.8                 & 235520       &  10240   & $0.77^{+0.04}_{-0.04}$ &  $0.50^{+0.09}_{-0.09}$  \\
1s  &  100     & 4.8                 & 222720       & 23040   & $0.75^{+0.04}_{-0.04}$ &  $0.49^{+0.07}_{-0.07}$  \\
3  &  100     & 5.3             & 130560       &  5120  &  $0.91^{+0.03}_{-0.03}$  &  $0.89^{+0.04}_{-0.04}$\\
5a &  100     & 9.6                 & 120320       &  2560   & $0. 79^{+0.04}_{-0.04}$ & $0.59^{+0.08}_{-0.08}$\\
6 &  100     & 9.6                 & 120320       &   2560   & $0.77^{+0.04}_{-0.04}$ &  $0.52^{+0.08}_{-0.07}$\\
7  &  100     & 10.6           & 266240       &  5120   & $0.83^{+0.04}_{-0.04}$  &  $0.68^{+0.06}_{-0.06}$ \\
14a &  300     & 13.5              & 170240       &  2560   & $0.93^{+0.03}_{-0.03}$ &  $0.90^{+0.06}_{-0.05} $\\ 
17 &  300     & 15.3           & 193280       &   2560   & $0.84^{+0.05}_{-0.04}$ & $0.72^{+0.08}_{-0.08}$ \\
18 &  300     & 27.0                & 171520       &   1280  & $0.91^{+0.03}_{-0.03}$  & $0.86^{+0.05}_{-0.05}$ \\
19 &  300     & 27.0                & 171520       &   1280  & $0.86^{+0.05}_{-0.04}$ & $0.76^{+0.08}_{-0.08}$ \\
20 &  300     & 30.6           & 194560       &  1280   &  $0.89^{+0.05}_{-0.05}$ &  $0.81^{+0.06}_{-0.06} $ \\
21 &  300     & 30.6           & 194560       &   1280   &  $0.91^{+0.03}_{-0.03}$  & $0.85^{+0.05}_{-0.05}$\\
\hline
1.1  &  100     & 4.8                & 942080       & 40960  & $0.76^{+0.04}_{-0.04}$  &  $0.51^{+0.08}_{-0.08}$\\
1s.1 &  100     & 4.8                & 890880       & 92160  & $0.74^{+0.04}_{-0.04}$ &  $0.49^{+0.09}_{-0.09}$\\
3.1  &  100     & 5.3                & 1044480      & 40960  & $0.90^{+0.05}_{-0.05}$ &  $0.89^{+0.08}_{-0.08}$\\
17.1 &  300     & 15.3               & 773120       &  10240   & $0.84^{+0.04}_{-0.04}$ & $0.72^{+0.08}_{-0.09}$ \\
20.1 &  300     & 30.6               & 1556480      & 10240  & $0.90^{+0.05}_{-0.05}$ &  $0.83^{+0.06}_{-0.07} $ \\
\hline
\end{tabular}

\medskip

{First column: name of the simulation. $\rccl$:~initial cluster break
radius (kpc). $\Mcltot$:~cluster mass ($10^{13}\Msol$).  $\Ndm$:~total
number of DM particles.  $\Nlum$:~total number of stellar
particles. $\rdmfin/\rccl$:~best--fitting break radius of the final DM
profile normalized to the initial break radius. $\beta$:~best--fitting
logarithmic slope of the final DM profile. $1\sigma$ uncertainties on
$\rdmfin/\rccl$ and $\beta$ are reported. 
Simulations~1.1, 1s.1, 3.1, 17.1, and 20.1 are higher-resolution replicas of
simulations~1, 1s, 3, 17, and 20, respectively. Pairs of 
runs with the same cluster parameters (5a and 6, 18 and 19, 20
and 21) differ in the initial centre-of-mass positions and velocities 
of the galaxies. In simulations~1s and 1s.1 50 small galaxies are added. }
\end{table}

\subsection{Models}

For the cluster DM distribution we use a Hernquist (1990) model
\begin{equation}
\label{hden}
\rhocl (r)= {\Mcltot \rccl \over 2 \pi r (\rccl +r)^{3}},
\end{equation}
with $\Mcltot$ and $\rccl$ cluster total mass and break radius,
respectively.  Thus, $\rhocl (r) \sim r^{-1}$ for $r \to 0$, as in the
NFW density profile, but, unlike the NFW profile, the total mass is
finite, because $\rhocl (r) \sim r^{-4}$ for $r \to \infty$. In the
simulations the distribution function of the cluster has been chosen
isotropic or radially anisotropic (see N03); $\rccl$ is either $100\,
\kpc$ or $300 \,\kpc$, and the total cluster mass is in the range
$4.8\times 10^{13} -3.06 \times 10^{14} \Msol$ (see Table~1).

At the beginning of each simulation the five galaxies are identical
(isotropic or radially anisotropic) Hernquist models: their stellar
density distribution $\rhostar$ is described by equation (1), with
break radius $\rcstar \simeq 2.2$ kpc and total mass $\Mstar=4 \times
10^{11} \Msol$.  The initial centre-of-mass positions and velocities
of the five galaxies are chosen by extraction from the cluster
distribution function, constrained by the observational data (see
N03).  Note that simulations with the same cluster parameters (namely,
the pairs of simulations~5a and 6, 18 and 19, 20 and 21; see Table~1)
differ in the initial centre-of-mass positions and velocities of the
galaxies.  In simulations~1s and 1s.1 we consider an additional
population of 50 smaller galaxies (each modeled as a Hernquist model
with $\Mstar = 5 \times 10^{10}\Msol$ and $\rcstar \simeq 0.55$ kpc)
distributed in phase--space according to the cluster distribution
function (see Fig.~1).

The total time of each simulation is $\sim 6.1$ Gyr, of the order of
$100\,\Td$, where $\Td$ is the half--mass dynamical time of the
initial galaxy models (we adopt $\Omega_{\rm m}=0.3$, $\Omega_{\rm
\Lambda}=0.7$, and $H_0=65$ km s$^{-1}$ Mpc$^{-1}$).

\subsection{Codes and numerical tests}

For most of the numerical N-body simulations (the first 12 listed in
Table~1) we used the Springel, Yoshida \& White (2001) GADGET code,
with cell--opening parameter $\alpha=0.02$, minimum and the maximum
time step $\Dtmin=0$ and $\Dtmax =\Td/100$, time step tolerance
parameter $\alphatol=0.05$, and softening parameter
$\varepsilon\simeq0.36 \rcstar$.  Additional higher-resolution
simulations (1.1, 1s.1, 3.1, 17.1, 20.1) were obtained using the
parallel code FVFPS (Fortran Version of a Fast Poisson Solver;
Londrillo, Nipoti \& Ciotti 2003).  This code, based on algorithms
introduced by Dehnen (2002), has been successfully tested and compared
with GADGET (Nipoti, Londrillo, Ciotti 2003a; Nipoti 2003). In
particular, we adopted the following values of the parameters: minimum
value of the opening parameter $\thetamin=0.5$; softening parameter
$\varepsilon\simeq0.1\rcstar$; initial time step $\Dt \sim \Td/100$.
In all simulations stellar and DM particles have the same mass, and
the total energy is conserved within $1 \%$.

\begin{figure}
\begin{center}
\parbox{1cm}
{
\psfig{file=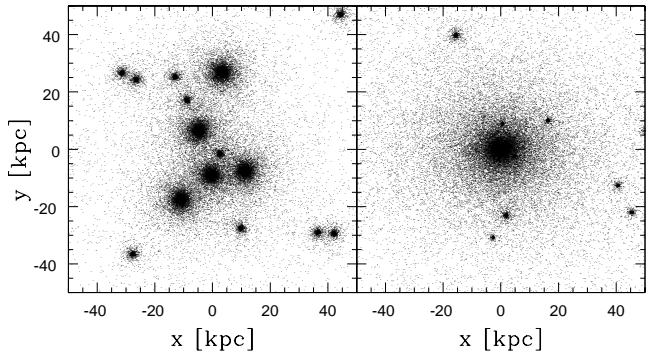,width=0.5\textwidth,angle=0,bbllx=18bp,bblly=144bp,bburx=591bp,bbury=455bp,clip=}
} 
\caption{Simulation~1s.1: snapshots of the initial ($z=0.59$, left)
and final ($z=0$, right) projected distribution of the stellar
particles in the inner cluster region.}
\end{center}
\end{figure}

Discreteness, two-body relaxation and softening are expected to
artificially alter the very inner DM density profile.  To quantify
numerical effects, we ran a few test simulations where the initial
conditions are represented by the cluster DM halo only, for the
adopted codes and values of the parameters (softening, number of DM
particles, time step, opening parameter, total elapsed time). The
final profile of one of these test simulations (with the same
parameters as simulation~1.1) is shown in Fig.~2 (empty symbols).  In
all the test simulations the Hernquist profile is preserved for $r
\gsim 0.05 \rccl$. Thus, we only consider the properties of the
cluster DM halo for $r \gsim 0.05 \rccl$ in the analysis of all the
presented simulations. We note that in the initial conditions of the
lowest-resolution of our simulations $r \sim 0.05 \rccl$ contains
$\sim 270$ particles. Thus, our choice of the minimum reliable radius
is consistent with recent numerical convergence studies (Power et al.
2003, and references therein).
  
Other possible numerical artifacts might be due to the small number of
particles used to represent each galaxy (128 to 2048, in the
simulations run with GADGET). In particular, it is likely that in our
simulations we overestimate the galaxy mass that is tidally stripped
(see Kazantzidis et al. 2004) and this might affect the response of
the cluster DM. To explore this problem we ran simulations~1.1, 1s.1,
3.1, 17.1, and 20.1, which have the same initial conditions as
simulations~1, 1s, 3, 17, and 20, respectively, but higher total
number of particles ($\sim 10^6$ DM particles, and 2048 to 8192
stellar particles for each of the five galaxies). We checked that all
the relevant properties of the system evolve in the same way in the
corresponding higher and lower resolution simulations. In particular
the final cluster DM profile is practically indistinguishable in the
two cases over the considered radial range $r \gsim 0.05 \rccl$.

\begin{figure}
\begin{center}
\parbox{1cm}{ \psfig{file=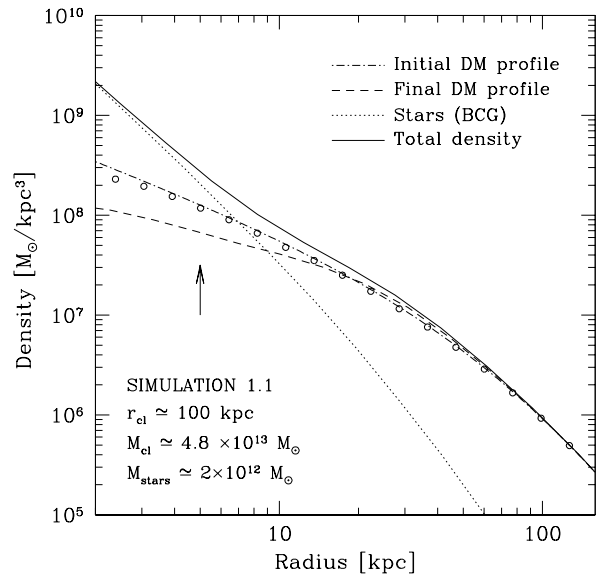,width=0.45\textwidth}}
\caption{Final density distribution of the cluster dark, luminous and
total matter in simulation~1.1. The dot-dashed curve is the initial
(Hernquist) DM profile.  The arrow indicates the minimum radius
considered for the fit. Empty symbols refer to a test simulation run
to exclude numerical artifacts (see Section~2.3).}
\end{center}
\end{figure}

\section{Results}

In all the simulations 3 to 5 out of the 5 massive galaxies merge
before $z=0$ at the bottom of the cluster potential well, as a
consequence of dynamical friction. Fig.~1 shows, as an example, the
initial ($z=0.59$, left) and final ($z=0$, right) projected
distributions of the {\it stellar} particles in the inner $\sim 50\,
\kpc$ for simulation~1s.1. In any case, the merger remnant is similar
in its main structural and dynamical properties to a real BCG (see
N03).  Here we discuss the effects of this dissipationless
multiple merging event on the underlying cluster DM distribution.

\subsection{Properties of the final dark matter density profile}

To quantify the combined effects of dynamical friction heating and
dissipationless contraction, we fit the final angle-averaged DM
density profile using a generalization of the Hernquist profile, in
the form (Dehnen 1993)
\begin{equation}
\label{genhden}
\rhodm (r)= {\rho_{\rm DM,0}  \rdm^4 \over r^{\beta} (\rdm +r)^{4-\beta}},
\end{equation}
where the inner slope $\beta$ and the break radius $r_{\rm DM}$ are
free parameters, and the reference density $\rho_{\rm DM,0}$ is
constrained by the total DM mass. The initial cluster DM density
profile (equation~\ref{hden}) corresponds to the case $\beta=1$,
$r_{\rm DM}=\rccl$, and $\rho_{\rm DM,0}=\Mcltot / (2 \pi\rccl^3)$.
 When computing the angle-averaged density profile of the final DM
  distribution, we determine the centre of the cluster using the
  iterative technique described by Power et al.  (2003): the centre of
  mass of the system is computed recursively, considering particles
  within a sphere whose radius shrinks by 2.5 per cent at each step.
  We stop the iteration when the sphere contains $\sim 1000$
  particles\footnote{As a check, we also computed the centre of
    the system as the position of the lowest-potential energy
    particle, finding good agreement with the adopted method.}.  The
best--fitting $\beta$ and $\rdm$ for the final cluster DM distribution
of each simulation are reported in Table~1; $1\sigma$ uncertainties on
the best--fitting parameters are calculated from $\Delta \chi^2=2.30$
contours in the space $\beta-\rdm$.

Fig.~2 shows the final dark, luminous and total matter distribution
for simulation~1.1: the final cluster DM distribution (dashed line) is
centrally shallower than the initial Hernquist profile (dot-dashed
line). The modification is apparent at $r \lsim 20$ kpc, where the
contribution of the newly formed central galaxy (dotted line) to the
total density (solid line) becomes relevant.  Thus, {\it the central
cluster DM cusp after the formation of the BCG is flatter than the
original $\rhodm \propto r^{-1}$}.

We find the same qualitative behaviour of the final luminous and dark
matter distributions in all the other simulations, though
simulation~1.1 represents one of the cases where flattening is more
effective (the final best--fitting $\beta$ is $\sim 0.5$). Considering
all of our simulations, we find best--fitting logarithmic slope
$\beta$ in the range $0.49 \lsim \beta \lsim 0.90$, with average
$\avbeta=0.71$.  Comparing the best--fitting final break radius
$\rdmfin$ with the break radius $\rccl$ of the corresponding initial
Hernquist model, we find $0.74\lsim \rdmfin/ \rccl\lsim0.95$, with
average $\avrdmfin=0.85$.  We note that the {\it best--fitting} final
break radius is not a direct measure of the concentration of the
system, because it can be significantly affected by the change in the
shape of the density distribution. In fact, the final radii containing
10, 50, 70 and 90 per cent of the total DM mass are always larger than
95 per cent of the corresponding initial values, even for quite small
$\rdmfin / \rccl$ (e.g., simulations~1 and 1s). Thus, in our models,
deviations of $\rdmfin$ from $\rccl$ indicate a difference in the
shape of the profile, as well as the value of $\beta$ does.
  
Simulation~1s.1 has the same initial conditions (cluster mass, radius
and distribution function, and phase space coordinates of the centres
of mass of the galaxies) as simulation~1.1, but in the first case 50
smaller galaxies are added to the five massive galaxies to represent
the cluster population (see Fig.~1).  We find that the main properties
of the final matter distribution in simulation~1s.1 are not
significantly different from those of simulation~1.1. We conclude that
even a rather simple model, representing only the more massive
galaxies in the initial conditions, is sufficient to capture the main
physical processes.

Thus, in the considered scenario we find that the most important
effect of the formation of a BCG through galactic cannibalism on the
underlying diffuse DM distribution is to produce a shallower final
cluster DM distribution. In other words, dynamical friction heating is
more effective than contraction due to the mass infall.

\subsection{Physical interpretation}

\begin{figure}
\begin{center}
\parbox{1cm}{ \psfig{file=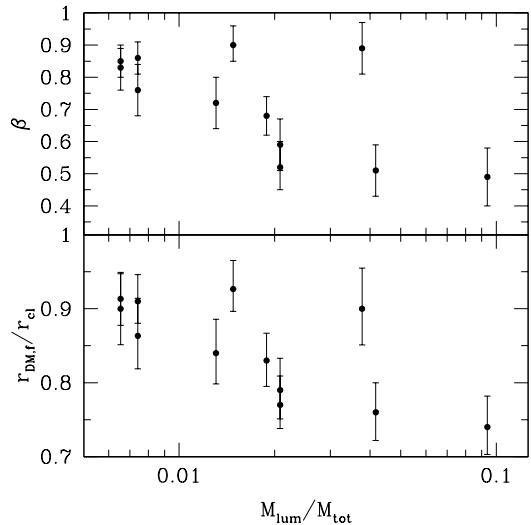,width=0.4\textwidth}}
\caption{ Best--fitting logarithmic inner slope $\beta$ (top) and
  break radius $\rdmfin$ (normalized to the initial break radius;
  bottom) of the final DM density distribution versus the ratio
  between the total mass in stars and total cluster mass (DM plus
  stars). Bars indicate $1\sigma$ uncertainties.}
\end{center}
\end{figure}

In Fig.~3 we plot the final best--fitting $\beta$ (top) and
$\rdmfin/\rccl$ (bottom) as a function of the ratio $M_{\rm
  lum}/\Mcltot$ between the total mass in stars and the total cluster
mass (DM plus stars).  The general trend is that the final density
profile is shallower when $M_{\rm lum}/\Mcltot$ is higher.  However,
the distribution of points in the plane $M_{\rm lum}/\Mcltot$-$\beta$
(as well as in the plane$M_{\rm lum}/\Mcltot$-$\rdmfin/\rccl$) is
characterized by a significant scatter: for example, simulations~1 and
3 have roughly the same $M_{\rm lum}/\Mcltot \sim 0.04$, but
significantly different final inner slopes ($\beta\sim 0.5$ and $\beta
\sim 0.9$, respectively).  This suggests that the cluster mass is not
a discriminant in determining the final density profile.  We argue
that the orbital properties of the five galaxies represent the main
factor affecting the final slope of the DM cusp. This picture would be
consistent with our findings, because simulations with the same or
similar $M_{\rm lum}/\Mcltot$ (and cluster break radius) differ only
in the initial phase-space coordinates of the galaxies in the cluster.
In general, galaxies on different orbits will heat the underlying DM
at different radii and at different rates. Thus, it is not surprising
to find a range of values of $\beta$ in simulations with similar
cluster parameters, but different initial positions and velocities of
the galaxies. 

In order to quantify this effect, we present here the results of a
very simple exercise. We ran two additional test simulations, with the
same number of particles, cluster parameters and initial positions of
the five galaxies as simulation~1, but different initial
velocities. In one case each galaxy starts with half the initial
kinetic energy it has in simulation~1 (and the same direction of the
velocity vector), in the other the galaxies have null initial
velocities. Interestingly, we find best--fitting final slopes
$\beta\simeq0.68$ in the former case, and $\beta\simeq1.29$ in the
latter, to be compared with $\beta\simeq 0.50$ for simulation~1. In
other words, for fixed initial position of a galaxy, the smaller is
its initial kinetic energy, the smaller the amount of energy it
transfers to the cluster DM through dynamical friction heating.  We
also recall that the infall of the galaxies into the cluster centre
deepens the cluster gravitational potential well.  If the amount of
dynamical friction heating is quite small (as in the extreme case of
the test simulation in which the galaxies start at rest), then
dissipationless contraction dominates and the final cusp is steeper
than the initial.  

We conclude that the orbital parameters of the galaxies are important
in determining the final DM distribution, as clearly illustrated by
the results of the two test simulations presented above.  However, it
must be stressed that these test simulations -- intended only to
isolate an important physical effect -- are characterized by quite
artificial initial conditions, which are not extracted from a
distribution function and are not consistent with the observational
phase--space constraints considered in our work (see N03).

\subsection{Comparison with observations}

How do the results of our simulations compare with observationally
determined cluster DM density profiles?  The galaxy clusters observed
by Sand et al. (2004) are suitable for this kind of exercise.  These
clusters (MS2137-23, Abell 383, RXJ1133, Abell 963, MACS1206, Abell
1201) are at redshifts lower than that of C0337-2522, they have total
mass of the order of a few $10^{14} \Msol$, and they host a dominant
giant elliptical (with mass around $10^{12}\Msol$) apparently relaxed
at the bottom of the cluster potential well. Sand et al. (2004) found
slopes $0.38 \lsim \beta \lsim 0.99$ with an average value of $\avbeta
\sim 0.52$ for the first three clusters, and an upper limit $\beta <
0.57$ for the others.  It appears that our simulations are able to
reproduce their observations.  In particular, it is very suggestive to
compare the final matter distribution of simulation~1.1 with the
best--fitting model that Sand et al. found for MS2137-23 (cf. Fig.~2
with their Fig.~7). The similarity of the DM and stellar distributions
is striking: the best--fitting slope is $\beta=0.57$ for MS2137-23 and
$\beta=0.50$ for the end--product of simulation 1.1. Not only can the
proposed mechanism reconcile the shallow observed slopes with the
steep slopes obtained in DM only simulations, but also the various
realizations of the cannibalism process could introduce additional
scatter, thus helping to explain the observed scatter.

Ultimately, when more observations are available and we can measure
the {\it distribution} of the final slopes it might be possible to use
the observed scatter to constrain the history of cannibalism in
clusters.

\section{Summary and conclusions}

In this paper we explored, with the aid of N-body simulations, the
effects on the inner cluster DM density profile of the formation of a
BCG, through dissipationless multiple merging of pre-existing
galaxies. The initial conditions are designed to reproduce the
observed galaxy cluster C0337-2522 ($z=0.59$). Although we focus on a
single well observed system as a case study, we stress that the
results of the simulations could be generalized to poor galaxy
clusters where the central giant elliptical was formed as a
consequence of galactic cannibalism, with negligible dissipation.

We found that the inner cluster DM profile is sensitive to the
dynamical evolution of cluster galaxies, and in particular to the
formation of a BCG.  In our models, where the initial condition are
realistic and the galaxies are deformable, we find final slopes in the
range $0.49 \lsim \beta \lsim 0.90$, with average $\avbeta=0.71$.  The
qualitative behaviour of our simulations is in agreement with what
found by El-Zant et al. (2004), though they obtained a quite small
value of the final inner slope ($\beta \simeq 0.35$), which is likely
to be due to the use of rigid clumps, combined with a high fraction of
luminous (clumped) matter ($M_{\rm lum}/\Mcltot \simeq 0.06$).
Dynamical friction heating in a cluster with original inner density
profile $\rhodm \propto r^{-1}$ can thus produce a flatter final inner
slope for the cluster DM. Recent measures of the DM profiles of
observed clusters can be interpreted taking into account this
mechanism.

As outlined in the Introduction, our results are based on the
assumption that cluster galaxies are essentially made of stars.
However, in a scenario in which the contribution of galactic DM haloes
to galaxy masses is important, the inspiraling galaxies might bring a
significant amount of DM in the centre of the cluster.  For instance,
if we consider, as extreme case, that the massive galaxies in our
simulations are made of DM only, the final DM cusp would be steeper
than $r^{-1}$, because the DM infall is dominant over dynamical
friction heating (cf. solid curve in Fig.~2).  A similar scenario has
been recently considered by Ma \& Boylan-Kolchin (2004), who studied
the interplay between a DM halo and its sub-haloes (see also Zhang et
al. 2002).  Ma \& Boylan-Kolchin show with N-body simulations that the
dynamical evolution of sub-haloes, depending on their mass and
concentration, can either steepen or flatten the halo cusp. Thus, the
determination of the amount and distribution of DM in cluster galaxies
is very relevant also to the problem of the cluster cusp.

As a consequence, the discrepancies found between the observationally
determined inner DM profiles and the predictions of CDM models must
not necessarily represent a failure of CDM, at least at the cluster
scale.  Instead, they could reflect our poor understanding of the
formation and evolution of galaxies, where the physics of baryons is
important.  These processes have to be taken into account when
predicting the properties of DM haloes of clusters, on scales of few
tens of kiloparsecs or smaller. At the same time, our simulations --
when interpreted in the context of structure formation (i.e., galaxies
considered as sub-haloes of DM) -- indicate that is quite remarkable
to obtain a universal DM profile as a result of successive
aggregations\footnote{We note that a non--homology effect of
  successive dissipationless mergings was already reported by Nipoti
  et al. (2003a).}, due to the requirement that dynamical friction
heating, potential well contraction, and mass increase all balance.
This detailed balance should be further studied in cosmological
simulations.

\section*{Acknowledgments}

We are grateful to Giuseppe~Bertin, James~Binney, Jeremiah~Ostriker,
David~Sand, Graham~P.~Smith, and the anonymous referee for helpful
comments on the draft. TT also acknowledges support from NASA through
Hubble Fellowship grant HF--01167.01.



\bsp

\label{lastpage}

\end{document}